# PerfXplain: Debugging MapReduce Job Performance

Nodira Khoussainova, Magdalena Balazinska, and Dan Suciu
Department of Computer Science and Engineering,
University of Washington, Seattle, WA, USA

{nodira, magda, suciu}@cs.washington.edu

## ABSTRACT

While users today have access to many tools that assist in performing large scale data analysis tasks, understanding the performance characteristics of their parallel computations, such as MapReduce jobs, remains difficult. We present PerfXplain, a system that enables users to ask questions about the relative performances (i.e., runtimes) of pairs of MapReduce jobs. PerfXplain provides a new query language for articulating performance queries and an algorithm for generating explanations from a log of past MapReduce job executions. We formally define the notion of an explanation together with three metrics, relevance, precision, and generality, that measure explanation quality. We present the explanation-generation algorithm based on techniques related to decision-tree building. We evaluate the approach on a log of past executions on Amazon EC2, and show that our approach can generate quality explanations, outperforming two naïve explanation-generation methods.

## 1. INTRODUCTION

The volume of data collected by businesses today is rapidly increasing. This data includes web crawls, search logs, click streams, network monitoring logs, and others. At the same time, tools for analyzing that data are becoming increasingly powerful and easy to use. Examples include parallel database management systems [27, 39, 47], MapReduce-based systems [14, 19, 28, 31, 40, 41], and others [9, 13, 33, 46, 52]. Large-scale clusters for carrying out this analysis are also becoming common-place. As a result, vast volumes of data are analyzed every day by a large variety of users.

Increasingly, users who write the MapReduce programs [19, 28], Pig Latin scripts [40], or declarative queries (e.g., HiveQL [31] or SQL) to analyze the data are not experts in parallel data processing, but are experts in some other domain. They need to ask a variety of questions on their data and these questions keep changing. For these users to be successful, they need to be self-sufficient in their data analysis endeavours. They cannot rely on administrators or distributed systems experts to help them debug and tune their analysis workloads, because there simply are not enough experts.

While most users already have tools to test and debug the correctness of their SQL queries or MapReduce programs before running them at massive scale, there are limited tools to help understand, diagnose, and debug any performance problems. The performance of parallel programs can be challenging to understand. As an example, when a user runs a MapReduce job and the job seems to take an abnormally long time, the user has no easy way of knowing if the problem is coming from the cluster (e.g., high load or machine failures), from some configuration parameters, from the job itself, or from the input data.

In this paper, we present PerfXplain, a system that assists users in *debugging the performance of MapReduce applications in a shared-nothing cluster*. PerfXplain lets users formulate performance queries in its own language called the PerfXplain Query Language (PXQL). A PXQL query identifies two MapReduce jobs or tasks. Given the pair of jobs (tasks), the query can inquire about their relative performances: e.g., Why did two MapReduce jobs take the same amount of time even though the second one processed half the data? Why was the last task in a MapReduce job faster than any of the other tasks in that job?

Given a query in PXQL, PerfXplain automatically generates an *explanation* for this query. Informally, an explanation consists of two predicates that hold true about the pair of identified executions. The first predicate, which we refer to as the *despite clause*, maximizes the probability of seeing the expected behavior. Meanwhile, the second predicate, called the *because clause*, maximizes the probability of the observed behavior. For example, if a user asks "why was the last task in this MapReduce job faster than any of the other tasks", an explanation might be: "even though the last task processed the same amount of data as the other tasks (despite clause), it was faster most likely because the overall memory utilization on the machine was lower (because clause) when it executed". When the predicate in the despite clause is true, a pair of tasks typically has the same runtime. *Within that context*, the because clause then explains why the user observed a performance different than anticipated. The despite clause thus helps ensure that the explanation given by the because clause is *relevant* to the identified pair of tasks, rather than just producing a generally-valid argument.

Hence, unlike prior work, which focused on predicting relational query performance [25, 26], predicting MapReduce job performance [24, 37, 38], automatically tuning MapReduce jobs [10, 20, 29, 30, 34] or relational queries [6, 8, 15, 16], and automatically diagnosing failures [21], the goal of PerfXplain is to *explain* the performance similarity or difference between pairs of MapReduce job or task executions. In this paper, we focus on explaining runtimes, but our approach can directly be applied to other performance metrics. Additionally, while our implementation and evaluation focus on MapReduce jobs, PerfXplain represents the execution of a single job or task as a vector of features, where each configuration parameter and runtime metric is a feature. As such, the approach is more broadly applicable.

PerfXplain uses machine learning to generate explanations. All performance queries in PerfXplain take the following form: The user specifies what behavior he or she *expected* (e.g., "I expected the last task to take the same amount of time as the others"), optionally why the user expected that behavior (e.g., "all tasks executed the same join algorithm"), and what behavior the user *observed* (e.g., "the last task was faster than the others"). To produce its explanations, PerfXplain utilizes a log of past MapReduce job executions along with their detailed configuration and performance metrics. Given a PXQL query, PerfXplain, identifies positive examples (pairs of jobs/tasks that performed as the user expected), and negative examples (pairs of jobs/tasks that performed as the user observed). From these examples, PerfXplain learns both the most likely reason why the pair should have performed as expected and, within that context, the most likely cause why the pair performed as observed. PerfXplain generates explanations from these two models. The key challenge for generating these explanations is to ensure that every explanation is highly precise, and at the same time as general as possible so that the user can apply this newly acquired knowledge to other scenarios. Overall, we make the following contributions:

1. We propose a simple language, PXQL, for articulating queries about the performance of a pair of MapReduce jobs or tasks (Sections 3.1 and 3.2).

2. We formally define the notion of a *performance explanation* and three metrics *relevance*, *precision*, and *generality* to assess the quality of an explanation (Section 3.3).

3. We develop an approach for *efficiently* extracting performance explanations that have high relevance, high precision, and good generality from a log of past MapReduce job executions (Section 4).

4. We evaluate the approach using a log of MapReduce jobs executed on Amazon EC2 [1]. We show that PerfXplain is able to generate explanations with higher precision than two naïve explanation-generation techniques, and offer a better trade-off between precision and generality (Section 6).

## 2. MOTIVATION AND OVERVIEW

We start with a motivating scenario that illustrates the need for PerfXplain. We then present the key types of performance queries that PerfXplain is designed to answer.

### 2.1 PerfXplain Motivation

Parallel data processing systems, such as MapReduce, can exhibit wildly varying performances when executing jobs. Indeed, the performance of a given MapReduce job depends on (1) the details of the computation to perform, (2) the volume of data that must be processed and its characteristics (such as the distribution of values in the input data, which can cause imbalance in processing times between tasks), (3) the current load, hardware configuration, and health of the cluster where the computation is being carried out, and (4) the configuration parameters for the cluster and for the job (block size, number of reducers, amount of memory allocated to the combiner [28], etc.).

Today, it is difficult for users to understand and fix any performance problems associated with their MapReduce computations. Working with scientists at the University of Washington, we have seen numerous cases of these problems. We have even faced such challenges ourselves.

As an example, consider a user who executes a MapReduce job on a 32GB dataset in a cluster with 150 machines. The job takes 30 minutes to run but produces a wrong answer. To debug her job, the user decides to execute it on a smaller, 1GB, dataset. By reducing the size of the dataset, the user hopes to speed-up her debug cycle. However, the smaller dataset also takes 30 minutes to run. Today, the user has limited tools to figure out why both datasets took the same amount of time to process, while the user expected a significant runtime improvement.

PerfXplain's goal is to help users debug this type of performance problem. In this case, the user would pose the following query:

I expected job $J_2$ to be much faster than job $J_1$. Why did it take the same amount of time to run?

In this scenario, the explanation is: "because the block size is large". Indeed, because the block size was set to a recommended value of 128 MB, the 32 GB dataset was split into 256 blocks and the 1 GB dataset was split into 8 blocks. Each machine can run two concurrent map and two concurrent reduce tasks (i.e., each machine has two map and two reduce slots), and thus neither the small nor large dataset used the full cluster capacity. The processing time was the time it takes to process one block of data, which is the same for both datasets.

Given such an explanation, the user can then take action. For example, she can reduce the block size or perhaps choose to debug the query locally on the 1GB dataset.

### 2.2 Types of Performance Queries

PerfXplain is designed to answer a variety of queries related to MapReduce application performance. Queries about runtimes refer to two MapReduce jobs or to two MapReduce tasks. The reasoning for this is that a user's expectation for how long a job should take, in general, comes from past experience. This is why we require the user to identify another job as a point of reference. Similarly, tasks have abnormal runtimes only in relation to the runtime of other tasks. By identifying the second job or task, the user clarifies where his runtime expectations come from. We identify two basic types of queries that users may have about the duration of a MapReduce job or task. The first type of queries ask why runtimes were different. The second type asks why runtimes were the same. We illustrate this classification with the following examples:

EXAMPLE 1. ***Different durations***. *I expected job $J_2$ to be much faster/slower than job $J_1$. However, they have almost the same durations. Why?*

EXAMPLE 2. ***Same durations***. *I expected job $J_1$ and $J_2$ to have a similar duration. However, $J_2$ was much faster/slower than $J_1$. Why?*

Additionally, performance queries can either be general queries as above or can be constrained queries by the addition of a *despite* clause. Constrained queries can help produce more relevant explanations as we demonstrate in Section 6.

EXAMPLE 3. ***Different durations (constrained)***. *Despite having less input data, job $J_2$ had the same runtime as $J_1$. I expected $J_2$ to be much faster. What is the explanation?*

EXAMPLE 4. ***Same durations (constrained)***. *Despite having a similar input data size and both using the same number of instances, $J_2$ was much slower than $J_1$. I expected both to have a similar duration. What is the explanation?*

Finally, similar types of queries can be asked both for jobs and for tasks. Task runtimes can be compared within and across jobs.

EXAMPLE 5. *I expected all map tasks to have similar durations since they processed the same amount of data. However, task $T_2$ was faster than the other tasks, e.g., $T_1$. Why was this the case?*

## 3. PERFORMANCE QUERIES

We introduce PerfXplain's data model and language.

### 3.1 Data Model

**Job and Task Representation:** To generate its explanations, PerfXplain assumes that it has access to a log of past MapReduce job executions. PerfXplain models job executions using the following schema for jobs:

Job(JobID,feature$_1$,...,feature$_k$,duration)

and the following schema for MapReduce tasks:

Task(TaskID,JobID,feature$_1$,...,feature$_l$,duration).

The features for MapReduce jobs include configuration parameters (e.g., DFS block size, number of reduce tasks), system performance metrics (e.g., metrics collected by Ganglia [2]), data characteristics (e.g., input data size), and application-level details (e.g., the relational operator corresponding to the MapReduce job if the job was generated from HiveQL or Pig Latin). In our current implementation, the features for tasks include all features that are collected in the MapReduce log files (e.g. the task type, map input bytes, map output bytes), the MapReduce job it belongs to, as well as all the system performance metrics collected by Ganglia during the task execution. PerfXplain comes configured with collecting these specific features but can easily be extended to use additional features.

Throughout the paper, we will refer to job and task executions with their JobID or TaskID, respectively. To refer to the value of a feature f for a specific job J, we will use the notation J.f.

**Representation of Examples:** Because PerfXplain answers queries about *pairs* of jobs (or tasks), all the examples that it learns from come in the form of pairs of jobs. We refer to a pair of jobs as a *training example*. A training example consists of $4 \cdot k$ features, where $k$ is the number of features that we collect for a single job or task (we call these the *raw features*).

Table 1 lists the features that we compute for each training example. The left column enumerates the set of features (which we refer to as F), and the right column specifies the domain for each feature. We assume that we know the domains of the raw features. We denote the domain of a feature f with dom(f).

The computed features (i.e., those listed in Table 1) encode the relationship between the two jobs for each raw feature, at varying levels of resolution. The first set of features, which are of the form f$_i$_isSame, are binary features that represent whether the two jobs have the same value for feature$_i$. The second set of features, of the form f$_i$_compare, represent whether $J_1$'s value for feature$_i$ is much less than (LT), similar to (SIM) [1], or much greater than (GT) $J_2$'s value for feature$_i$. This feature is appropriate only for numeric features and thus the value of the feature is set to be missing for nominal features. Similarly, the third set of features, which are of the form f$_i$_diff represent the change in value for feature$_i$. This feature is computed only for nominal features, and is thus set to be missing if a feature is numeric. For example, if the value for pigscript for J$_1$ is filter.pig and for J$_2$ is join.pig, then the value of pigscript_diff is (filter.pig, join.pig). We refer to these three sets of features as *comparison features*, because they compare the raw features of the two jobs (or tasks). Finally, the fourth set of features are directly copied from the jobs if the jobs have the same value for that feature. Namely, feature f$_i$ is set to the value J$_1$.feature$_i$ if J$_1$.feature$_i$ = J$_2$.feature$_i$. Otherwise, the feature is labeled as missing. We refer to these features as the *base features*.

---
[1] In the current implementation, two values are considered to be similar if they are within 10% of one another.

| Feature | Domain |
|---|---|
| f$_1$_isSame | {T, F} |
| ... | |
| f$_k$_isSame | {T, F} |
| f$_1$_compare | {LT, SIM, GT} |
| ... | |
| f$_k$_compare | {LT, SIM, GT} |
| f$_1$_diff | dom(feature$_1$) × dom(feature$_1$) |
| ... | |
| f$_k$_diff | dom(feature$_k$) × dom(feature$_k$) |
| f$_1$ | dom(feature$_1$) |
| ... | |
| f$_k$ | dom(feature$_k$) |

**Table 1: Set of features that define a training example. The features are computed for a pair of jobs (tasks), and encode the relationship between the two jobs (tasks) for each raw feature, at varying levels of resolution.**

The key intuition behind the above feature choice is that they span the range from general features (i.e., _isSame features) to specific features (i.e., base features). The general features help abstract details when they are not important, which has two implications. First, explanations can become more generally applicable. Second, pairs of jobs that have very different raw features can become comparable. For example, if a task had a different runtime than another because the load on the instance was different, PerfXplain can generate an explanation of the form "CPU utilization isSame = false" rather than "CPU utilization when running task 1 was X while CPU utilization when running task 2 was Y". At the same time, detailed features are sometimes needed to get precise explanations when details matter. For example, the reason why a job took the same amount of time as another even though it used more instances could be "because the block size was larger than or equal to 128MB".

### 3.2 PXQL Syntax

PXQL allows users to formulate queries over the performance of either MapReduce jobs or tasks. To simplify the presentation, we focus only on jobs in this section.

A PXQL query consists of a pair of jobs and three predicates over their features. The first two predicates describe the observed behavior for the two jobs and the reason why the user is surprised by this behavior. The third predicate specifies what behavior the user expected. Every predicate takes the form $\phi_1 \land \ldots \land \phi_m$, where each $\phi_i$ is of the form $f$ op $c$ where $f$ is a feature from Table 1, $c$ is a constant, and op is an operator. The set of operators supported by PerfXplain include $=, \neq, <, \leq, >$ and $, \geq$.

DEFINITION 1. *A* **PXQL query Q** *comprises a pair of jobs* $(J_1, J_2)$ *and a triple of predicates* (**des**, **obs**, **exp**), *where* **des**, **obs** *and* **exp** *are predicates over $J_1$ and $J_2$'s features. Additionally,* **des**$(J_1, J_2)$ = true, **obs**$(J_1, J_2)$ = true, *but* **exp**$(J_1, J_2)$ = false. *Furthermore, it must be the case that* **obs** $\vDash \neg$**exp**.

*We refer to* $(J_1, J_2)$ *as the* **pair of interest**, *and the predicates as the* **despite**, **observed**, *and* **expected** *clauses, respectively.*

```
1. OBSERVED duration_compare = SIM
   EXPECTED duration_compare = GT
2. OBSERVED duration_compare = LT
   EXPECTED duration_compare = SIM
3. DESPITE inputsize_compare = GT
   OBSERVED duration_compare = SIM
   EXPECTED duration_compare = GT
4. DESPITE inputsize_compare = SIM ∧
           numinstances_isSame = T
   OBSERVED duration_compare = LT
   EXPECTED duration_compare = SIM
5. DESPITE inputsize_compare = SIM ∧
           jobID_isSame = T
   OBSERVED duration_compare = GT
   EXPECTED duration_compare = SIM
```

**Figure 1: Example PXQL queries.**

We use the following syntax for PXQL queries.

```
FOR J1, J2 WHERE J1.JobID = ? and J2.JOBID = ?
DESPITE des OBSERVED obs
EXPECTED exp
```

Informally, a PXQL query $Q = (\mathbf{des}, \mathbf{obs}, \mathbf{exp})$ over the pair of jobs $J_1, J_2$ can be read as "Given jobs $J_1$ and $J_2$, despite **des**, I observed **obs**. I expected **exp**. Why?" PerfXplain's goal is to then reply with an explanation of the form:

```
DESPITE des'
BECAUSE bec
```

where $\mathbf{des'}$ is a an extension of the user's **despite** clause and **bec** is a predicate over the features of the MapReduce jobs that appeared in the query.

Figure 1 shows how each example from Section 2 translates into a PXQL query. We omit the FOR clause. For example, the first query asks why the two jobs had a similar duration (duration_compare = SIM) and that the user expected that $J_1$ would be slower than $J_2$ (duration_compare = GT). As illustrated in the first two examples, the **despite** clause is optional. Omitting the clause is equivalent to setting **des** to $true$. Example 5 shows that the same query language that we use for jobs serves to ask performance queries over tasks.

### 3.3 PXQL Semantics

Given a PXQL query, PerfXplain must present the user with an explanation.

DEFINITION 2. *For a query $Q = (\mathbf{des}, \mathbf{obs}, \mathbf{exp})$ over a pair of jobs $(J_1, J_2)$, a* **candidate explanation** *$E$ is a pair of predicates $(\mathbf{des'}, \mathbf{bec})$. The predicates are referred to as the* **despite**, *and* **because** *clauses, respectively.*

For instance, for Example 1, a candidate explanation is $E = (\mathbf{des'}, \mathbf{bec})$ where $\mathbf{des'} = $ (inputsize_compare = GT) and $\mathbf{bec} = $ (blocksize >= 128 MB ∧ numinstances ≥ 100).

The first requirement from an explanation is that it holds true for the pair of jobs that the user is asking about. For example, explanation $E$ above says that the reason why the durations of $J_1$ and $J_2$ were similar is because the two jobs both had a large block size and a large number of instances. However, this explanation would not make sense if $J_1$ and $J_2$ did not satisfy these conditions. In such a case, we say that $E$ is not applicable to $(J_1, J_2)$.

DEFINITION 3. *A candidate explanation $E = (\mathbf{des'}, \mathbf{bec})$ is* **applicable** *to a pair of jobs $(J_1, J_2)$ if $\mathbf{des'}(J_1, J_2) = $ true and $\mathbf{bec}(J_1, J_2) = $ true.*

The applicability requirement for an explanation is a hard requirement. Every explanation generated by PerfXplain *must* be applicable. Additionally, we define three metrics of the quality of an explanation for a given log of MapReduce job executions.

DEFINITION 4. *The* **relevance**, $Rel(E)$, *of an explanation $E = (\mathbf{des'}, \mathbf{bec})$ given a PXQL query $(\mathbf{des}, \mathbf{obs}, \mathbf{exp})$ is the following conditional probability:*

$$Rel(E) = P(\mathbf{exp}|\mathbf{des'} \wedge \mathbf{des}). \quad (1)$$

Intuitively, an explanation with high relevance identifies (through the $\mathbf{des'} \wedge \mathbf{des}$ clause) the key reasons why the pair of jobs *should have* performed as expected. For example, if we consider our explanation $E$ from above, it has a high relevance because its **des** clause specifies that it consider only pairs of jobs where inputsize_compare = GT. Indeed, given that the input size of $J_1$ is greater than $J_2$, we would expect that $J_1$ be slower than $J_2$. By considering only pairs of jobs that satisfy the $\mathbf{des'} \wedge \mathbf{des}$ clause, the explanation given by the **bec** clause is more relevant because it focuses on circumstances that are specific to the user query. In our example, the **bec** clause identifies why pairs of jobs where one job consumes a much greater input still can have the same runtime. This explanation is more relevant to the query than one which would have explained why a job can have the same runtime as another job, *in general*.

DEFINITION 5. *The* **precision**, $Pr(E)$, *of an explanation $E = (\mathbf{des'}, \mathbf{bec})$ given a PXQL query $(\mathbf{des}, \mathbf{obs}, \mathbf{exp})$ is the following conditional probability:*

$$Pr(E) = P(\mathbf{obs}|\mathbf{bec} \wedge \mathbf{des'} \wedge \mathbf{des}). \quad (2)$$

A precise explanation tries to identify why, in the context of $\mathbf{des'}$ and **des**, did the pair in question most likely perform as it did instead of as expected. For example, consider $E' = (\mathbf{des'}, \mathbf{bec'})$ where $\mathbf{bec'} = $ blocksize >= 128 MB. This is a shorter version of $E$ from above. $E'$ has most likely a lower precision than $E$ because it is rarely the case that two jobs have a similar runtime just because they have the same large block size. On the other hand, if the two jobs also executed in a large cluster, then it is likely that neither used the full cluster capacity and the runtime was determined by the time to process one large block of data.

Though precision is necessary, an explanation with high precision may still be undesirable. Consider the following **because** clause: start_time = 1323158533 ∧ instance_url = 12-31-39-E6.compute-1.internal:localhost/127.0.0.1. Such an explanation can have a precision of 1.0, yet it is still not a good explanation. A good explanation is one that can apply to more than one setting. In fact, we posit that the more settings where an explanation applies, the better the explanation, because it identifies more general patterns in job performance. We measure this third property with the following metric.

DEFINITION 6. *The* **generality**, $Gen(E)$, *of an explanation $E = (\mathbf{des'}, \mathbf{bec})$ given a PXQL query $(\mathbf{des}, \mathbf{obs}, \mathbf{exp})$ is the following conditional probability:*

$$Gen(E) = P(\mathbf{bec}|\mathbf{des'} \wedge \mathbf{des}). \quad (3)$$

Note that precision and generality are closely related to the data mining concepts of confidence and support, respectively. The only difference is that our terms explicitly refer to the various clauses of the explanation. Namely, precision is the confidence that the **because** clause leads to observed behavior in the context of the **despite** clause, and generality is the support of the **because** clause in the context of the **despite** clause.

Given a PXQL query, PerfXplain's goal is to generate an applicable explanation that achieves high precision, relevance and generality. However, as in the example above, precision and generality are usually in direct conflict with one another. Thus, a helpful explanation must strike a good balance between the two metrics.

Finally, PerfXplain orders the predicates in the **despite** and **because** clauses so that the important predicates appear first. A predicate is more important than another if it achieves higher marginal relevance (in the **despite** clause) or higher marginal precision (in the **because** clause).

## 4. PXQL QUERY EVALUATION

In this section, we describe how PerfXplain generates explanations for PXQL queries. We begin with a few definitions.

### 4.1 Terminology

Given a PXQL query and a pair of jobs in the log, we first say that the pair of jobs is *related* to a query if it satisfies the **des** clause and either the **expected** or **observed** clauses.

DEFINITION 7. *A pair of jobs* $(J_1, J_2)$ *is* **related** *to a PXQL query* $Q = (\textbf{des}, \textbf{obs}, \textbf{exp})$ *if* $\textbf{des}(J_1, J_2) = \text{true} \land (\textbf{exp}(J_i, J_j) = true \lor \textbf{obs}(J_i, J_j) = true)$.

Further, we say that a related pair of jobs performed as expected or as observed with respect to the query depending on whether it satisfied the **expected** or **observed** clause. More formally:

DEFINITION 8. *A pair of jobs* $(J_i, J_j)$ **performed as expected** *with respect to a PXQL query* $Q = (\textbf{des}, \textbf{obs}, \textbf{exp})$ *if* $\textbf{des}(J_i, J_j) = \text{true} \land \textbf{exp}(J_i, J_j) = true$.

Similarly,

DEFINITION 9. *A pair of jobs* $(J_i, J_j)$ **performed as observed** *with respect to a PXQL query* $Q = (\textbf{des}, \textbf{obs}, \textbf{exp})$ *if* $\textbf{des}(J_i, J_j) = \text{true} \land \textbf{obs}(J_i, J_j) = true$.

### 4.2 Approach

Given a query $Q$, PerfXplain generates an explanation in the form of a pair of **des**' and **bec** clauses. The constructions of these two clauses is symmetrical. We first explain how PerfXplain generates the **bec** clause.

**Overview of bec clause generation.** The **bec** clause generation takes two inputs. The first input is the log of past MapReduce job executions. Each pair of jobs in the log forms a training example, which is represented by a combined vector of features as shown in Table 1. These job pairs and their features serve as the basis for generating the explanation. The second input is the PXQL query itself. The query comprises the pair of jobs of interest, $(J_1, J_2)$ and the three predicates: (**des**, **obs**, **exp**).

The key idea behind performance explanation is to identify the conditions why the pair of interest performed as observed rather than performing as expected. This condition takes the form of a predicate on the job-pair features (i.e., those listed in Table 1). As discussed in the previous section, we want an explanation that is both precise and general: an explanation is precise if whenever a pair of jobs satisfies it, that pair is likely to perform as observed. At the same time, an explanation is general if it applies to many pairs of jobs in the log.

**Detailed algorithm for bec clause generation.** Algorithm 1 shows the detailed **bec** clause generation approach. The algorithm takes as input a PXQL query, the pair of interest $(J_1, J_2)$, the set of all jobs $J$, and the desired explanation width $w$. The width is the number of atomic predicates in the explanation.

*Lines 1-2: Construct training examples.* The first step in the explanation generation process (i.e., constructTrainingExamples) identifies the related pairs of jobs in the log. Only pairs that satisfy the **des** predicate and either the **obs** or **exp** predicates are used to generate an explanation for the given query. The **obs** and **exp** predicates also serve to classify job pairs as performing either as observed or as expected. Next, the algorithm keeps just a sample of this set. We further discuss sampling in Section 4.3.

Given these training examples, the algorithm generates the explanation as a conjunction of atomic predicates. It grows the explanation by adding atomic predicates in a greedy fashion. To select each atomic predicate, the algorithm identifies (a) the "best" predicate for each feature, and then (b) selects the "best" predicate across features.

*Line 5: Construct best predicate for each feature.* An atomic predicate is of the form $f$ op $c$. Thus, given a feature $f$, in order to find the best predicate, PerfXplain must select the best op and constant $c$ pair. For nominal attributes, the only operator it considers is equality. For numeric attributes, it considers both equality and inequality operators. In order to select the best predicate for a feature, PerfXplain identifies the predicate with the highest information gain, which is defined as:

$$Information\_Gain(P, \phi) = H(P) - H(P|\phi)$$

where $\phi$ is the predicate, $P$ is the pairs of jobs in consideration, and $H(P)$ is the information entropy of $P$. When we consider $\phi$, we think about the two partitions that $\phi$ creates: the pairs that satisfy $\phi$ and the pairs that do not. By maximizing the information gain, we want to find the predicate that leads to two partitions where the partitions each have a lower entropy (or higher "purity") than the entropy of the full set of pairs.

As an example, consider Figure 2. The leftmost box represents the full set of training examples we are considering. The training examples are depicted with either a + (if an example performed as observed) or with a − (if it performed as expected). Suppose for a feature $f$, we are considering two predicates. For example, for blocksize we may be considering blocksize > 64MB and blocksize ≤ 256MB. The two predicates are illustrated by the second (A) and third (B) boxes in Figure 2. The grey area represents where a training example satisfies the predicate. Now, if we consider the two predicates, clearly $A$ is a better predicate than $B$, because it is doing a better job of separating the +'s from the −'s. The information gain metric captures exactly this intuition. Entropy is defined as $H(P) = -plog_2(p) - (1-p)log_2(1-p)$ where $p$ is the fraction of +'s. In our example, $p = 0.6$ for the full sample. Thus, our original entropy is 0.97. The entropy of the sample in $A$ is 0.1, which is calculated by taking the weighted average of the entropies of both the grey side and the white side (the two partitions created by the predicate). Thus, the information gain for $A$ is 0.87. For $B$, the entropy is 0.97, which is an information gain of 0. Therefore, the predicate depicted in $A$ is better than the predicate depicted in $B$.

*Lines 6-15: Identify the best cross-feature predicate.* For each of the above per-feature predicates, the algorithm computes its precision and generality. Both precision and generality are measured over the set of job-pairs ($P$ in Algorithm 1) that are related to the PXQL query and satisfy the explanation constructed so far. We compute precision as the number of jobs-pairs that satisfy the predicate and perform as observed, divided by the number of job-pairs in that satisfy the predicate. Generality is the fraction that satisfies the predicate. The score of a predicate then becomes a weighted average of its precision and its generality scores (line 13). In the current implementation, we use a weight of $w = 0.8$ (thus favoring precision over generality).

Note, however, that the score is not simply a weighted average over the raw precision and generality scores. Instead, it calculates a relative score for each. Consider the precision score, or $precScore$, as an example. To calculate it, PerfXplain computes the precisions of all the predicates, ranks them, and replaces the precision values with the percentile ranks. PerfXplain does the same transformation for the generality score. In our earlier implementation, we had not included this step, and we found that because the generality scores tended to be much lower than the precision scores (especially as the explanation grew in width), the generality was not having enough impact on the predicate score. Therefore, we introduced this step to normalize the two scores before taking their weighted average. Finally, the predicate with the highest score is added to the explanation.

*Lines 16-18: Extend explanation and continue.* To further refine the explanation, PerfXplain iterates and adds additional atomic predicates. At each iteration, PerfXplain considers only those job pairs that satisfy the **bec** predicate generated so far. Some of the job pairs still performed as expected in that set. PerfXplain then identifies an additional atomic predicate that isolates these job pairs from the others while correctly classifying the pair of interest. The resulting extended predicate forms a more precise explanation for why the pair of jobs performed as observed rather than performing as expected. The algorithm stops once a clause of width $w$ has been generated.

The result of the algorithm is an explanation consisting of a **bec** clause with $w$ atomic predicates.

**Generating the des' clause.** Given a PXQL query, the **bec** predicate strives to capture a precise yet general reason why some jobs performed as observed rather than performing as expected. The **bec** predicate is restricted to hold for the pair of interest identified in the query. In spite of this constraint, we found that it was often the case that the explanation would produce *overly generic* reasons why a pair of jobs performed as observed rather than performing as expected. For example, consider the case where a user asks why two jobs had the same runtime instead of one job being faster than the other. In the absence of a **des** clause, a general and precise explanation of width 1 says that the two jobs executed on the same number of instances. Instead, if the system generates the explanation not by using the entire log but by only considering the subset of job pairs where one job processed a significantly larger amount of data than the other, the most precise and general explanation changes. For this subset of jobs, the explanation becomes about block size and cluster size. The latter explanation is more relevant to the pair of interest. The **des** clause captures this intuition in a principled fashion.

In the current implementation, by default, PerfXplain generates only the **bec** clause in an explanation, and the user must explicitly request a **des** clause. An easy modification is to set a relevance threshold $r$. If the user's **des** clause achieves a relevance score less than $r$, then PerfXplain will extend the clause automatically until its score reaches this threshold or it can not further be improved.

Conveniently, the **des**' clause generation technique is symmetric to the **bec** clause generation. In order to generate the **des**' clause, PerfXplain uses the same algorithm as shown in Algorithm 1. How-

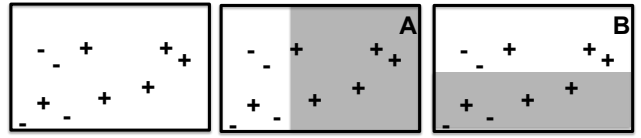

**Figure 2:** Example of information gain.

**Algorithm 1** PerfXplain Algorithm

**Input:** PXQL query $q = (\mathbf{des}, \mathbf{obs}, \mathbf{exp})$, jobs $J_1$ and $J_2$, set of all jobs $J$, width $w$
**Output:** an explanation X
1: $P \leftarrow constructTrainingExamples(J, q)$
2: $P \leftarrow sample(P, J_1, J_2)$
3: $X \leftarrow true$
4: **for** $i = 1 \ldots w$ **do**
5: $\quad predicates \leftarrow [\, maxInfoGainPredicate(f) : f \in F \,]$
6: $\quad precisions \leftarrow [\, P(obs|p, X) : p \in predicates \,]$
7: $\quad generalities \leftarrow [\, P(p|X) : p \in predicates \,]$
8: $\quad predScores \leftarrow []$
9: $\quad$ **for** $j \in [1 \ldots |predicates|]$ **do**
10: $\quad\quad p \leftarrow predicates[j]$
11: $\quad\quad precScore \leftarrow normalizeScore(precisions[j], precisions)$
12: $\quad\quad genScore \leftarrow normalizeScore(generalities[j], generalities)$
13: $\quad\quad predScores.append((p, w \cdot precScore + (1 - w) \cdot genScore)$
14: $\quad$ **end for**
15: $\quad (bestPred, bestScore) \leftarrow \text{argmax}_{p \in predScores}\, p.score$
16: $\quad X \leftarrow X \wedge bestPred$
17: $\quad P \leftarrow [\, p : p \in P \wedge X \text{ holds true for } p\,]$
18: **end for**
19: **return** $X$

ever, it changes line 6 to measure relevance $P(\mathbf{exp}|p)$ instead of precision $P(\mathbf{obs}|p)$.

Once PerfXplain has generated a sufficient **des'** clause, PerfXplain verifies the clause with the user. If the user approves this clause, it is added to the user's PXQL query, and PerfXplain can proceed to generating the **bec** clause, and thus a full explanation.

**Comparison to other machine-learning techniques.** Explanation generation is related to classification problems in machine learning. In particular, our approach is related to decision trees [42] since both identify predicates over features that separate examples into two classes (observed and expected in our case). There are however several important distinctions. First, unlike a decision tree, performance explanation must ensure that the pair of interest is always correctly classified as performing as observed. Second, performance explanation need not categorize all pairs of jobs in the log. Instead, it must generate a predicate that yields a relevant, precise, and general explanation *given the pair of interest*. In order to achieve this goal, performance explanation must construct a **des'** clause *before* generating the **bec** clause. Additionally, it must consider the precision and generality metrics during the construction of each of these two clauses.

While we cannot apply decision trees directly to the performance explanation problem, we still re-use the notion of information gain for constructing the best predicate for each feature. In our prototype, we use the C4.5 [42] technique for finding the predicate that maximizes the information gain metric.

### 4.3 Sampling

To maintain a low response-time for the explanation generation, PerfXplain samples the training examples related to the current query (line 2 of Algorithm 1). Sampling also helps balance the number of positive and negative examples that will contribute to the explanation. A balanced sample is one in which there is ap-

603

proximately the same number of examples labeled as *observed* and as *expected*. A highly unbalanced sample can cause PerfXplain to believe that a trivial explanation is sufficient. For example, if a sample consists of pairs where 99% performed as observed, PerfXplain will decide that the empty explanation is good as it will achieve a precision of 99%.

The sampling method operates as follows. It iterates through each training example. If the desired sample size is $m$ and $T$ is the set of all training examples, then the sampling technique keeps a training example $t$ with probability:

$$p = \begin{cases} m/(2 \cdot |\{x \in T : \mathbf{obs}(x) = true\}|) & \text{if } \mathbf{obs}(t) = true \\ m/(2 \cdot |\{x \in T : \mathbf{exp}(x) = true\}|) & \text{if } \mathbf{exp}(t) = true \end{cases}$$

For instance, consider a set of training examples that consists mostly of pairs labeled with *observed* and very few with *expected*. In this case, a training example labeled with *observed* will have a lower probability of being selected for the sample than a training example labeled with *expected*. In our current implementation, we use a sample size of 2000.

Currently, PerfXplain randomly samples training examples, which already yields high-quality explanations as we show in Section 6. Biasing the samples in some way, such as ensuring that priority is given to executions that correspond to a varied set of jobs, could possibly improve explanation quality further. We leave this question open for future work.

## 5. ALTERNATIVE APPROACHES

In this section, we describe two naïve techniques for constructing explanations. Though at first glance, both techniques seem like they may be sufficient for generating good explanations, we see that both fall short in different ways. We compare the PerfXplain approach to these techniques in Section 6.

### 5.1 RuleOfThumb Approach

This technique first identifies which features of a job have a high impact on the runtime of a job in general. Then it points to differences in these features as the explanation. This identification of important features is executed only once, and is not performed per PXQL query. Once the user issues a PXQL query along with a pair of interest $(J_1, J_2)$, RuleOfThumb returns the top-$w$ features that the two jobs disagree on (where $w$ is the width of the explanation desired). The features are ranked by their importance as determined in the step described above.

Consider the following example. Suppose that the initial stage identifies that numinstances, inputsize, and num_reduce_tasks are the most important features for determining the duration of a job, respectively. Suppose the user has asked why job $J_1$ is slower than job $J_2$, and that both jobs agree on the number of instances, but disagree on the input data size and on the number of reduce tasks. In this case, the explanation generated would be: inputsize_isSame = F ∧ num_reduce_tasks_isSame = F.

Any standard feature selection algorithm can be used to determine the most important features. In our implementation, we use the Relief technique [44] because it is able to handle both numeric and nominal attributes, as well as missing values.

The RuleOfThumb algorithm works well for some PXQL queries. For example, it may be appropriate for queries that ask for an explanation of why the runtime of two jobs is different because the technique always points to differences in important features and differences in features usually lead to differences in the runtime. However, this approach completely ignores the PXQL query, and will therefore, fail to satisfactorily answer many queries.

**Algorithm 2** SimButDiff Algorithm

**Input:** PXQL query $q = (\mathbf{des}, \mathbf{obs}, \mathbf{exp})$, jobs $J_1$ and $J_2$, set of all jobs $J$, width $w$, similarity threshold $s$
**Output:** an explanation X
1: $T \leftarrow constructTrainingExamples(J, q)$
2: $isSameFeatures \leftarrow [\, f : f \in F \wedge f \text{ is a }\_isSame feature\,]$
3: $T \leftarrow reduceDimensionality(T, isSameFeatures)$
4: $k \leftarrow s \cdot dimensionality(T)$
5: $S \leftarrow [\, t : t \in T \wedge t \text{ agrees with } (J_1, J_2) \text{ on } \geq k \text{ features}\,]$
6: $featureScores = [\,]$
7: **for** $f \in isSameFeatures$ **do**
8:    $d \leftarrow |[\, t : t \in S \wedge t.f \neq (J_1, J_2).f\,]|$
9:    $o \leftarrow |[\, t : t \in S \wedge t.f \neq (J_1, J_2).f \wedge \mathbf{exp}(t) = true\,]|$
10:    $featureScores.append((f, \frac{o}{d}))$
11: **end for**
12: $featureScores \leftarrow sort(featureScores)$
13: $X \leftarrow true$
14: **for** $i = 1 \ldots w$ **do**
15:    $(f, score) \leftarrow featureScores[i]$
16:    $X \leftarrow X \wedge (f = (J_1, J_2).f)$
17: **end for**
18: **return** $X$

### 5.2 SimButDiff Approach

Unlike the previous technique, the SimButDiff algorithm actually considers the PXQL query when generating its explanation. It first finds all training examples that are similar to the pair of interest, with respect to its _isSame features. Among these similar pairs, for each feature $f_i$, it measures the fraction of pairs that performed as expected and disagreed on this feature to the number of pairs that disagreed on this feature. In essence, it performs 'what-if' analysis on each feature $f_i$ to check the following: if this feature had been different, how likely is it that the pair would have performed as expected. For example, if the pair of interest agree on numinstances, it finds all pairs that were similar to the pair of interest, but disagreed on the numinstances to see if that generally leads to pairs that performed as expected. It measures this fraction for each feature, and the features that have the highest fractions constitute the explanation.

Algorithm 2 shows the details of this approach. In addition to the PXQL query, the pair of interest, the set of all jobs, and the desired width, the algorithm also takes as input a similarity threshold $s$ between 0 and 1. In the current implementation, a similarity threshold of 0.9 has worked well.

The algorithm proceeds as follows. First, like the PerfXplain algorithm, it constructs the training examples (line 1). However, it keeps only the _isSame features (lines 2-3). Next, it filters out training examples that are not similar to $(J_1, J_2)$, the pair of interest (lines 4-5). A training example is similar if it agrees with the pair of interest on at least $s$ fraction of the _isSame features.

Next, the algorithm iterates through every feature and calculates a score for it (lines 6-11). The score for a feature $f$ is the fraction of training examples that perform as expected among those that *disagree* with $(J_1, J_2)$ on $f$. The features are then sorted in descending order of these scores (line 12) and the explanation is a conjunction of predicates of the form $f = (J_1, J_2).f$, constructed in order of the score (lines 14-17).

The SimButDiff algorithm only utilizes the _isSame features because they are binary features, and thus it is easy to identify the training examples that disagree with $(J_1, J_2)$ on a feature (i.e., check that the training example takes on the one different value). Secondly, it is simple to measure the similarity of two training examples by just counting the number of features that they agree on. Were we to leverage some of the other features, we would need to define similarity scores between the different values in the domain of the feature.

| Parameter | Different Values |
|---|---|
| Number of instances | 1, 2, 4, 8, 16 |
| Input file size | 1.3 GB, 2.6 GB |
| DFS block size | 64 MB, 256 MB, 1024 MB |
| Reduce tasks factor | 1.0, 1.5, 2.0 |
| IO sort factor | 10, 50, 100 |
| Pig script | simple-filter.pig, simple-groupby.pig |

**Table 2: The parameters we varied and the different values for each.**

As such, because the SimButDiff technique leverages only the _isSame features, it fails to produce precise explanations where more complex features are required. We show examples of such PXQL queries in Section 6.

## 6. EVALUATION

In this section, we compare the three explanation generation approaches on two PXQL queries over real data that we collected on Amazon EC2. We explore several aspects of the techniques. In Section 6.3, we assume that the user has given us a well-specified PXQL query, and we compare the precisions of the explanations generated by each technique. In Section 6.4, we assume that the user has provided an under-specified query, and investigate whether PerfXplain is able to generate an effective **despite** clause. In Section 6.5, we explore the case where the log consists of only one type of MapReduce job, whereas the pair of jobs in question are of a different type. We evaluate the impact of the log size on the precision of explanations in Section 6.6. Finally, in Section 6.7, we analyze the trade-off between generality and precision. We begin with a description of the experimental setup (Section 6.1) and the two PXQL queries that we use for this evaluation (Section 6.2).

### 6.1 Experimental Setup

To collect real data for our experiments, we ran two different Pig scripts on Amazon EC2 and varied several parameters for each execution. Table 2 shows the features that we varied and the different values that we tried for each one.

The number of instances is the number of virtual machines used for the job. The input file consists of a log of search queries submitted to the Excite [23] search engine. This is a sample file that is used in the standard Pig tutorial. We concatenate the sample file from the tutorial to itself either 30 or 60 times. This input data file is then broken up into a given number of blocks. The block size is set through the dfs.block.size parameter in the MapReduce configuration file. The block size determines the number of map tasks that are generated for an input file (i.e., the number of map tasks is the input file size divided by the block size). The reduce tasks factor determines the number of reduce tasks for the job, using the mapred.reduce.tasks parameter. The factor is in relation to the number of instances. e.g.. If there are 8 instances, and the reduce tasks factor is 1.5, then the number of reduce tasks is set to 12. The IO sort factor feature corresponds to the io.sort.factor MapReduce parameter, and represents the number of segments on disk to be merged together at a given time. Finally, the Pig script parameter specifies which Pig job should be executed. The simple-filter.pig script simply loads the input file, filters out all queries where the query string is a URL, and outputs the queries that remain. The simple-groupby.pig script groups all the queries by user and outputs the number of queries per user.

PerfXplain collects data at the job-level as well as the task-level. For each task, PerfXplain extracts all details it can from the MapReduce log file, including hdfs_bytes_written, hdfs_bytes_read, sorttime, shuffletime, taskfinishtime, and tracker_name. We refer the reader to a Hadoop guide for details of these properties [49]. Additionally, PerfXplain also monitors the instances using Ganglia [2], which is a distributed monitoring system. Ganglia metrics include boottime, bytes_in, bytes_out, cpu_idle, and more. To be precise, PerfXplain runs Ganglia to measure these metrics on each instance once every five seconds. For a given task, it identifies the instance that the task was executed on, and for each metric, it calculates the average value while the task was executing. PerfXplain also percolates this monitoring data up to the jobs. Namely, for each job and each metric, it calculates the average value of the metric across all the tasks belonging to the job. In total, PerfXplain records a total of 64 features for each task and 36 features for each job.

The graphs in this section are generated as follows. We divide the log into two logs: the *training log* and the *test log*. This split is done randomly; we iterate through each job, add it to the training log with 50% probability, and all remaining jobs are added to the test log. The training log is used as the basis for generating the explanation. The test log is used to evaluate the explanation. Namely, we measure the precision of the explanation over the test log. We repeat this process ten times, and our graphs report the average results across these ten runs, along with errors bars to depict the standard deviation.[2]

### 6.2 The PXQL Queries

We evaluate how well PerfXplain generates explanations for two different PXQL queries. The first query asks why one task is faster than another, and the second asks why one job is slower than another. We measure the precision, relevance, and generality of the explanations generated.

Here are the two PXQL queries that we use:

1. **WhyLastTaskFaster:**
   FOR $T1, T2$
   DESPITE $jobID\_isSame = T \land$
   $inputsize\_compare = SIM \land$
   $hostname\_isSame = T$
   OBSERVED $duration\_compare = LT$
   EXPECTED $duration\_comare = SIM$

2. **WhySlowerDespiteSameNumInstances:**
   FOR $J1, J2$
   DESPITE $numinstances\_isSame = T \land$
   $pig\_script\_isSame = T$
   OBSERVED $duration\_compare = GT$
   EXPECTED $duration\_compare = SIM$

The first query asks why the last task on an instance runs faster than the earlier tasks that were executed on the same instance, even though each task processes a similar amount of data. Interestingly, we came across this query when we were puzzled by this pattern while collecting our experimental data. The reason we discovered was that the last task runs faster than earlier tasks because the instances have two cores and can run two tasks simultaneously. Sometimes, by the time the last task is reached, all other tasks are completed, and the instance is free to run only one task. Thus, the system load is lighter for the last task, and consequently the task is completed faster.

The second query asks why a job is slower than another job even though both jobs are running the same Pig script, and on the same number of instances. The explanation here is that the input size of the slower job is much greater than the input size of the faster job.

---

[2]A common evaluation method used in machine learning literature is ten-fold cross validation. We did not use this technique because it leads to a small test log consisting of a tenth of the jobs. A small log is ineffective for testing because it results in too few pairs that performed as observed. Therefore, we used 2-fold cross validation.

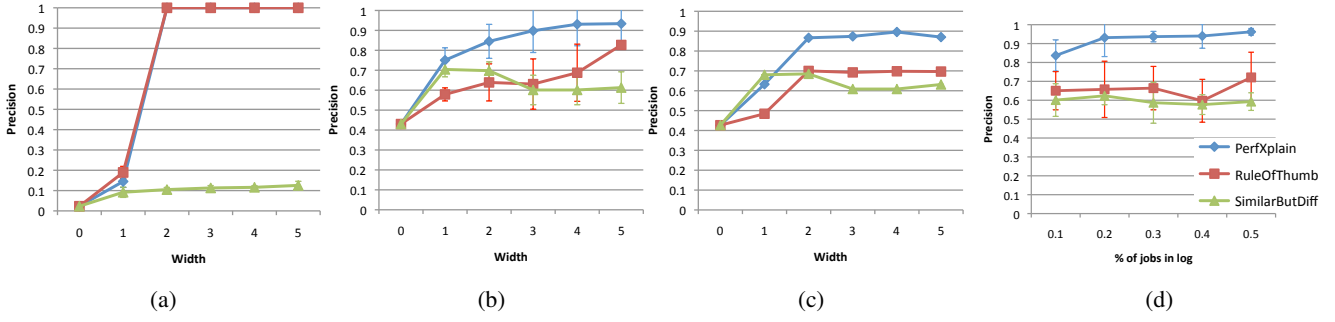

Figure 3: Explanation precisions for (a) WhyLastTaskFaster, and (b) WhySlowerDespiteSameNumInstances with varying width. Precisions for (c) when the log consists only of simple-groupby.pig jobs, and (d) at width-3 with varying log size for WhySlowerDespiteSameNumInstances.

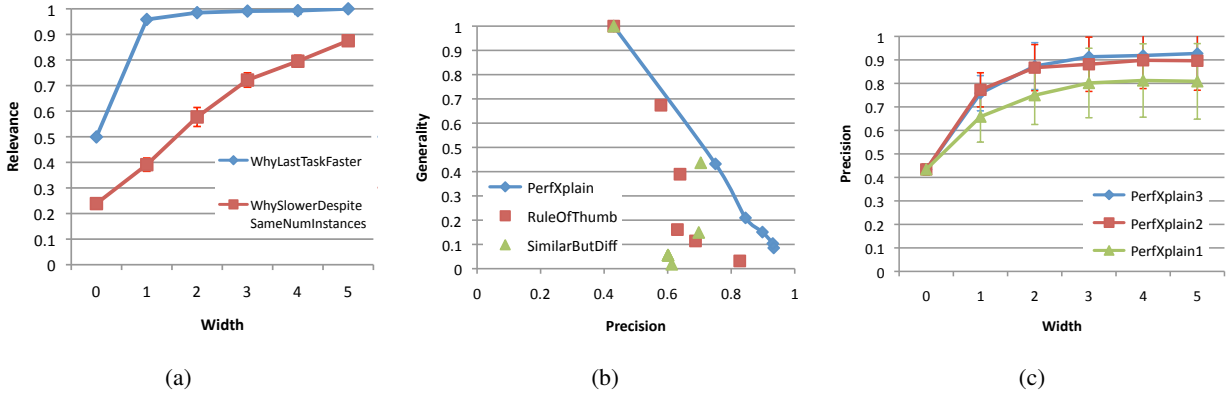

Figure 4: (a) Relevance of PerfXplain-generated despite clauses, (b) Precision versus generality for WhySlowerDespiteSameNumInstances, (c) Precision of explanations for WhySlowerDespiteSameNumInstances with different feature levels.

### 6.3 Well-specified PXQL Queries

In this section, we evaluate PerfXplain's explanation generation technique for PXQL queries where the user has specified a reasonable **des** clause as shown above. A good despite clause facilitates the generation of highly relevant explanations because the user manually constrains the search space. We compare the PerfXplain technique to the two naïve techniques described in Section 5.

Figure 3(a) shows the precision of the explanations generated by each technique for the **WhyLastTaskFaster** PXQL query. The x-axis indicates the width threshold specified for the explanation. Note that when the width is 0, the explanation is empty (or $true$) and thus the precision is $P(\mathbf{obs}|\mathbf{des} \wedge true) = P(\mathbf{obs}|\mathbf{des})$.

With the exception of one run, both the RuleOfThumb and PerfXplain approaches generate the same explanation (for width 3), and thus achieve similar precision: `avg_cpu_user_isSame = F ∧ avg_proc_total_isSame = F ∧ avg_load_one_isSame = F`. The explanation says the task was faster because the average CPU time spent on user processes is not the same, the average total number of processes is not the same, and the average load time across a minute is not the same. This explanation is generally pointing towards the fact that the system load is different for the two tasks, and thus this leads to the observed behavior. (In the one exceptional run, PerfXplain generates an explanation starting with `avg_load_five_isSame = F`, which can also lead to faster execution, but it achieves a slightly lower precision than the above explanation.) The SimButDiff technique generates explanation `avg_pkts_in_isSame = F ∧ avg_bytes_in_isSame = F ∧ avg_pkts_out_isSame = F`. The first part says that the average number of packets arriving is different, the second part talks about average number of bytes in, and so on. This explanation is well-grounded in the data; it is indeed the case that if two tasks have a similar number of packets arriving, they also are likely to perform as expected (i.e. have a similar duration). However, not many pairs of tasks have a similar number of packets arriving, and it is not the case that just because two tasks have a different number of packets arriving that they will achieve very different runtimes. The predicate `avg_cpu_user_isSame = F` appears in SimButDiff's explanation, but usually not until the seventh or eighth predicate.

Figure 3(b) shows the precision for the **WhySlowerDespiteSameNumInstances** task. Here, the PerfXplain approach generates the following explanation (for width 3): `inputsize_compare = GT ∧ avg_load_five_compare = GT ∧ numinstances <= 12`. The first predicate in the explanation indicates that the input size is larger, which is the correct explanation. It continues to explain that the average load (measured at five-minute intervals) for the slower job is higher, which is probably just a result of the larger input size. Finally, it says that the number of instances is small. This is also correct because if both jobs had a sufficiently high number of instances, the change in data size would not have affected the runtime.

The explanations of width 3 for RuleOfThumb and SimButDiff are `avg_load_five_isSame = F ∧ avg_proc_total_isSame = F ∧ inputsize_isSame = F` and `inputsize_isSame = F ∧ iosortfactor_isSame = T ∧ blocksize_isSame = T`, respectively. Though both techniques note that the input size has

606

| Query | Avg Relevance Before | Avg Relevance After |
|-------|---------------------|---------------------|
| 1 | 0.49 | 0.99 |
| 2 | 0.24 | 0.72 |

**Table 3: Relevance of PXQL queries with an empty despite clause versus with a PerfXplain-generated despite clause.**

an impact on the duration, they can only point to the fact that the input size is different (instead of greater than). Furthermore, the RuleOfThumb does not include `inputsize_isSame = F` until the third predicate because it is distracted by other side-effects of a larger input, which are that the average load and the total number of process is different.

*In summary, we see that PerfXplain generates explanations with a better or equal average precision than the two naïve techniques. For example, for the **WhySlowerDespiteSameNumInstances** query, at width 3, PerfXplain achieves at least 40.5% higher precision than both techniques.*

### 6.4 Under-specified PXQL Queries

In this section, we evaluate the quality of the **des** clauses generated by PerfXplain. We use the same PXQL queries as described in Section 6.2 but this time with the **des** clause removed. Table 3 above shows the relevance without the **des** clause, as well as the relevance with PerfXplain's automatically generated **des** clause. For this experiment, we restrict the clause to width 3.

Here are examples of **des** clauses that PerfXplain generates for each of the two queries:

1. `map_output_records_isSame = T ∧ tracker_name_isSame = T ∧ map_input_records_isSame = T ∧ file_bytes_written_isSame = T`

2. `pigscript_isSame = T ∧ numinstances_isSame = T ∧ blocksize_isSame = T`

For the **WhyLastTaskFaster** query (1), we see that the **des** clause indicates that the numbers of map output records and the number of input records are the same for both tasks, as is the name of the tracker. The second and third predicates are analogous to the user-specified **des** clause. The user-specified **des** clause achieved an average relevance of 0.97, whereas the PerfXplain-generated one achieves 0.99. For the second query, we see that PerfXplain generates exactly the **des** clause that the user specified with the additional `blocksize_isSame = T` predicate at the end. Thus, it achieves a slightly higher relevance score of 0.72 (the user-specified **des** clause had a relevance of 0.6).

Figure 4(a) shows the relevance score for both PXQL queries for **des** widths ranging from 0 to 5. Once again, width 0 represents the relevance of the empty **des** clause. We see that for both PXQL queries, PerfXplain is able to generate **despite** clauses with high relevance.

*In summary, we see that PerfXplain is able to generate a good* **despite** *clause if the user fails to do so, thus increasing the relevance of an empty* **despite** *clause by up to 200%.*

### 6.5 Explaining a Different Job

In this section, we explore whether the techniques can support a scenario in which the pair of interest is different from all the jobs in the log. This experiment is trying to answer the question: Can we use the approach to explain the performance of new jobs, different from those executed in the past? We analyze this scenario for the **WhySlowerDespiteSameNumInstances** query. The pair of interest for this PXQL query consists of two jobs that are both running the same Pig script: simple-filter.pig. The log, however, consists only of the data for the simple-groupby.pig jobs (plus the pair of interest).

In this experiment, we execute the three explanation-generation algorithms over the log described above, and evaluate the explanation precision over a log consisting of all the simple-filter.pig jobs. Figure 3(c) presents the results.

Comparing the results to those in Figure 3(b), we see that PerfXplain performs slightly worse than when it has access to a normal job log. For width-1 explanations, the precision is significantly lower when PerfXplain has access to only simple-groupby.pig jobs (0.63) versus with the full log (0.93). However, by width-3, the difference shrinks to 0.02 (from 0.89 to 0.87). The average decrease in precision across the different widths for PerfXplain is 0.04. SimButDiff performs almost equivalently across the two scenarios, achieving a slightly lower precision (average of 0.001 lower). The average decrease in precision for the RuleOfThumb technique is 0.02.

*In summary, we see that the precision of PerfXplain's explanations decrease slightly if the log consists of jobs that are different from the jobs in question. For example, for width-3 explanations we saw an average of only 2.7% decrease in precision.*

### 6.6 Varying the Log Size

We investigate the effect of the log size on the different techniques. Namely, we randomly selected $x\%$ of the jobs in the log to use as the training log and varied $x$ between 10% and 50%.

Figure 3(d) summarizes the results of the experiment for the **WhySlowerDespiteSameNumInstances** query. The x-axis represents the size of the log we used, and the y-axis reports the precision. The results shown are for width-3 explanations. We see that for the PerfXplain technique, the precision increases gradually with the size of the job log. However, even with just 10% of the log, PerfXplain achieves an average precision of 0.84. However, the standard deviation of the precision at 10% is much higher than at 50% (0.08 versus 0.02). In contrast, the precisions of the RuleOfThumb and SimButDiff techniques are not significantly impacted by the sample size. In fact, the SimButDiff approach does not seem to be impacted by the sample size at all. The RuleOfThumb approach is affected by the size of the log, and the general trend is that it generates better explanations when the log is larger (with the exception of 0.4). However, the variance of its precision is high.

*A key takeaway from this experiment is that even with a small query log consisting of only 10% of the jobs, PerfXplain is able to achieve a high precision of 0.84 for width-3 explanations.*

### 6.7 Precision and Generality Trade-off

As we discussed in Section 3.3, an effective explanation generation approach should achieve a good trade-off between precision and generality. Figure 4(b) plots the precision and generality scores of explanations generated by the different techniques. As the figure shows, PerfXplain achieves a better trade-off between generality and precision than the other approaches because its points fall higher and more to the right than the points for the other two approaches. (We connect PerfXplain's points to better show the positions of all its points in relation to the other techniques' points.)

### 6.8 Using Different Features

Using the right set of features is crucial to generating good explanations. On one hand, simpler features result in explanations that are generally applicable. On the other hand, having access to more complex features and just more features in general can lead to explanations that achieve higher precision.

In this section, we investigate the effect of different feature sets on the precision of the explanations that PerfXplain generates. We consider three different feature sets, we call them levels.

1. Level 1 includes only the _isSame features.
2. Level 2 includes the _isSame features, the _compare features, and the _diff features.
3. Level 3 includes the _isSame features, the _compare features, the _diff features, and the base features.

Figure 4(d) shows the precision of explanations for the **WhyS-lowerDespiteSameNumInstances** PXQL query at each feature level. As the figure shows, PerfXplain achieves a similar precision for both levels 2 and 3, which outperform level 1 by a significant margin. We see an improvement in precision for level 3 versus level 2, at width 3. Remember, that the explanation generated by PerfXplain for this approach is inputsize_compare = GT ∧ avg_load_five_compare = GT ∧ numinstances <= 12. The third predicate here says that it is because the number of instances is small. This feature is not included at Level 2, and thus we see the improvement at width 3.

*In summary, we see that if we limit the set of features to only the _isSame features, PerfXplain suffers significantly in terms of precision. However, hierarchy levels 2 and 3 perform similarly in this scenario.*

## 7. RELATED WORK

**Database performance tuning.** Many existing database management systems provide tools to examine and tune the performance of SQL queries including Teradata [11], Oracle [18], SQL Server [7, 8, 17], MySQL [3], DB2 [32], Postgres [4, 5], and others. These tools focus on tuning the physical and logical designs of a database. In contrast, our goal is to explain performance issues across all layers of the system: from input data through configuration parameters. Existing tools are also aimed at database administrators. They show the physical plan of a query and point-out potential problematic aspects such as table scans (instead of indexes), full table-to-table product joins, etc. In contrast, our goal is to help non-expert users understand the performance problems of their queries. Our explanations cannot require detailed understanding of how the system works. Finally, existing systems focus on the performance of queries that are issued by applications and are thus repeated frequently while we need to explain the performance of a MapReduce job that, perhaps, has been executed only once.

More recent work has focused on MapReduce tuning and optimizations [29, 20, 35, 45, 34]. Many of these techniques address the problem from a different perspective from traditional database optimizers. Consider Xplus [29] as an example. Given a query, whereas existing database optimizers search for a plan first, and then execute the plan, Xplus tries to find a better plan by running a small set of plans proactively, collecting monitoring data from these runs, and iterating until it finds a better plan.

**Autonomic databases.** Another line of work strives to make database systems self-configurable, self-tuning, or autonomic, including MapReduce systems [10, 30]. A key project in this space is the Auto-Admin project, which aims to make DBMSs self-administrating [16, 15, 8, 6]. This includes automatically selecting indexes and materialized views [16, 8], and allowing a DBA to explore hypothetical configurations [15]. Research on autonomic databases includes work on automating failure diagnosis [21].

However, even when the processing engine can tune itself, performance problems can also come from the analysis code or input data. As a result, even completely self-tuning systems will need to be able to explain their performance to users.

**Applying machine-learning to system modeling.** There have been many applications of machine-learning techniques to various systems challenges such as predicting query performance [26], automatically monitoring and identifying system problems from console logs [51], root cause analysis of databases running on storage area networks [12], optimizing multicore systems [25], as well as parameter tuning in relational database systems [22, 48]. In contrast, in this work, our goal is to apply machine learning to the problem of explaining the performance variations between different executions of parallel computations in the form of MapReduce jobs or tasks.

**Machine learning tools.** There exist many tools that are commonly used for performing machine learning. For example, Weka [50] is a Java library that implements a collection of machine learning algorithms. Though not specifically for machine learning, MATLAB [36] and R [43] are two popular tools for data analysis, which often involves machine learning. Though they do not provide unified libraries, the two tools provide good support for matrices, vectors, and have a suite of visualizations that are helpful in the machine learning workflow.

PerfXplain differs from these tools in that it is not a generic tool for performing machine learning. Instead, it focuses only on explaining why two specific jobs did not perform as expected. As such, it requires little from the user: a PXQL query and a log of past job executions. It does not require the user to construct training data matrices nor implement machine learning algorithms. That said, integrating PerfXplain with such tools would probably prove incredibly helpful in the data analysis process.

## 8. CONCLUSION

In this paper, we addressed the problem of debugging performance problems in MapReduce computations. We presented PerfXplain, a system that enables users to ask comparative performance-related questions about either pairs of MapReduce jobs or pairs of MapReduce tasks. Our current implementation considers only queries over job or task runtimes but the approach can readily be applied to other performance metrics.

Given a performance query, PerfXplain uses a log of past MapReduce job executions to construct explanations in the form of predicates over job or task features. PerfXplain's key contributions include (1) a language for articulating performance-related queries, (2) a formal definition of a performance explanation together with three metrics, relevance, precision, and generality for explanation quality, and (3) an algorithm for generating high-quality explanations from the log of past executions.

Experiments on real MapReduce job executions on Amazon EC2 demonstrate that PerfXplain can indeed generate high-quality explanations, outperforming two naïve explanation-generation methods. While our focus in this paper has been on MapReduce jobs, because PerfXplain simply represents job or task executions as feature vectors, the approach has the potential to generalize to other parallel data processing systems.

## 9. ACKNOWLEDGMENTS

This work is partially supported by the NSF Grant IIS-0627585, the NSF CAREER award IIS-0845397, a Google Research Award, and an HP Labs Innovation Research Award.

The authors would also like to thank Greg Malewicz from Google, YongChul Kwon, Prasang Upadhyaya, and the reviewers for their helpful feedback.

## 10. REFERENCES

[1] Amazon EC2. http://aws.amazon.com/ec2/.
[2] Ganglia Monitoring System. http://www.ganglia.sourceforge.net.
[3] MySQL Query Analyzer. http://www.mysql.com/products/enterprise/query.html.
[4] PostgreSQL Tuning Wizard. http://pgfoundry.org/projects/pgtune.
[5] Tuning Your PostgreSQL Server. http://wiki.postgresql.org/wiki/Tuning_Your_PostgreSQL_Server.
[6] S. Agrawal, S. Chaudhuri, L. Kollar, A. Marathe, V. Narasayya, and M. Syamala. Database Tuning Advisor for Microsoft SQL Server demo. In *Proc. of the SIGMOD Conf.*, pages 930–932, 2005.
[7] S. Agrawal, S. Chaudhuri, L. Kollár, A. P. Marathe, V. R. Narasayya, and M. Syamala. Database Tuning Advisor for Microsoft SQL Server 2005. In *Proc. of the 30th VLDB Conf.*, pages 1110–1121, 2004.
[8] S. Agrawal, S. Chaudhuri, and V. R. Narasayya. Automated Selection of Materialized Views and Indexes in SQL Databases. In *Proc. of the 26th VLDB Conf.*, pages 496–505, 2000.
[9] ASTERIX: A Highly Scalable Parallel Platform for Semi-structured Data Management and Analysis. http://asterix.ics.uci.edu/.
[10] S. Babu. Towards Automatic Optimization of MapReduce Programs. In *Proc. of the 1st ACM symposium on Cloud computing (SOCC)*, pages 137–142, 2010.
[11] D. Becker and P. A. Barsch. Strike it Rich: Application Tuning Helps Companies Save Money through Query Optimization. Teradata magazine online. http://www.teradata.com/tdmo/v07n04/FactsAndFun/Services/StrikeItRich.aspx.
[12] N. Borisov, S. Uttamchandani, R. Routray, and A. Singh. Why Did My Query Slow Down? In *Proc. of the Fourth CIDR Conf.*, 2009.
[13] R. Chaiken, B. Jenkins, P.-A. Larson, B. Ramsey, D. Shakib, S. Weaver, and J. Zhou. SCOPE: Easy and Efficient Parallel Processing of Massive Data Sets. In *Proc. of the 34th VLDB Conf.*, pages 1265–1276, 2008.
[14] C. Chambers, A. Raniwala, F. Perry, S. Adams, R. R. Henry, R. Bradshaw, and N. Weizenbaum. FlumeJava: Easy, Efficient Data-parallel Pipelines. In *PLDI '10: Proceedings of the 2010 ACM SIGPLAN conference on Programming language design and implementation*, pages 363–375, 2010.
[15] S. Chaudhuri and V. Narasayya. AutoAdmin What-if Index Analysis Utility. *SIGMOD Rec.*, 27:367–378, June 1998.
[16] S. Chaudhuri and V. R. Narasayya. An Efficient Cost-Driven Index Selection Tool for Microsoft SQL Server. In *Proc. of the 23rd VLDB Conf.*, pages 146–155, 1997.
[17] S. Chaudhuri and V. R. Narasayya. Self-Tuning Database Systems: A Decade of Progress. In *Proc. of the 33rd VLDB Conf.*, pages 3–14, 2007.
[18] B. Dageville, D. Das, K. Dias, K. Yagoub, M. Zaït, and M. Ziauddin. Automatic SQL Tuning in Oracle 10g. In *Proc. of the 30th VLDB Conf.*, pages 1098–1109, 2004.
[19] J. Dean and S. Ghemawat. MapReduce: Simplified Data Processing on Large Clusters. In *Proc. of the 6th OSDI Symp.*, pages 137–149, 2004.
[20] J. Dittrich, J.-A. Quiané-Ruiz, A. Jindal, Y. Kargin, V. Setty, and J. Schad. Hadoop++: Making a Yellow Elephant Run Like a Cheetah (Without It Even Noticing). *Proc. VLDB Endow.*, 3:515–529, September 2010.
[21] S. Duan, S. Babu, and K. Munagala. Fa: A System for Automating Failure Diagnosis. *Proc. of the 25th ICDE Conf.*, pages 1012–1023, 2009.
[22] S. Duan, V. Thummala, and S. Babu. Tuning Database Configuration Parameters with iTuned. *PVLDB*, 2(1):1246–1257, 2009.
[23] My Excite. http://www.excite.com/.
[24] A. Ganapathi, Y. Chen, A. Fox, R. H. Katz, and D. A. Patterson. Statistics-driven Workload Modeling for the Cloud. In *SMDB*, pages 87–92, 2010.
[25] A. Ganapathi, K. Datta, A. Fox, and D. Patterson. A Case for Machine Learning to Optimize Multicore Performance. In *HotPar*, 2009.
[26] A. Ganapathi, H. Kuno, U. Dayal, J. Wiener, A. Fox, M. Jordan, and D. Patterson. Predicting Multiple Performance Metrics for Queries: Better Decisions Enabled by Machine Learning. In *Proc. of the 25th ICDE Conf.*, pages 592–603, 2009.
[27] Greenplum Database. http://www.greenplum.com/.
[28] Hadoop. http://hadoop.apache.org/.
[29] H. Herodotou and S. Babu. Xplus: a SQL-Tuning-Aware Query Optimizer. *Proc. VLDB Endow.*, 3:1149–1160, 2010.
[30] H. Herodotou, H. Lim, G. Luo, N. Borisov, L. Dong, F. B. Cetin, and S. Babu. Starfish: A Self-tuning System for Big Data Analytics. In *Proc. of the Fifth CIDR Conf.*, 2011.
[31] Hive. http://hadoop.apache.org/hive/.
[32] IBM. DB2 Performance Tuning using the DB2 Configuration Advisor. http://www.ibm.com/developerworks/data/library/techarticle/dm-0605shastry/index.html.
[33] M. Isard, M. Budiu, Y. Yu, A. Birrell, and D. Fetterly. Dryad: Distributed Data-Parallel Programs from Sequential Building Blocks. In *Proc. of the European Conference on Computer Systems (EuroSys)*, pages 59–72, 2007.
[34] E. Jahani, M. Cafarella, and C. Re. Automatic Optimization for MapReduce Programs. *PVLDB*, 4(6):385–396, 2011.
[35] D. Jiang, B. C. Ooi, L. Shi, and S. Wu. The Performance of MapReduce: An In-depth Study. *PVLDB*, 3:472–483, 2010.
[36] MATLAB. *version 7.10.0 (R2010a)*. The MathWorks Inc., 2010.
[37] K. Morton, M. Balazinska, and D. Grossman. ParaTimer: A Progress Indicator for MapReduce DAGs. In *Proc. of the SIGMOD Conf.*, pages 507–518, 2010.
[38] K. Morton, A. Friesen, M. Balazinska, and D. Grossman. Estimating the Progress of MapReduce Pipelines. In *Proc. of the 26th ICDE Conf.*, pages 681–684, 2010.
[39] Netezza, inc. http://www.netezza.com/.
[40] C. Olston, B. Reed, U. Srivastava, R. Kumar, and A. Tomkins. Pig Latin: a Not-So-Foreign Language for Data Processing. In *Proc. of the SIGMOD Conf.*, pages 1099–1110, 2008.
[41] R. Pike, S. Dorward, R. Griesemer, and S. Quinlan. Interpreting the Data: Parallel Analysis with Sawzall. *Scientific Programming*, 13(4), 2005.
[42] J. R. Quinlan. *C4.5: Programs for Machine Learning*. Morgan Kaufmann Publishers Inc., San Francisco, CA, USA, 1993.
[43] R Development Core Team. *R: A Language and Environment for Statistical Computing*. R Foundation for Statistical Computing, Vienna, Austria, 2011.
[44] M. Robnik-Sikonja and I. Kononenko. An Adaptation of Relief for Attribute Estimation in Regression. In D. H. Fisher, editor, *Fourteenth International Conference on Machine Learning*, pages 296–304. Morgan Kaufmann, 1997.
[45] J. Schad, J. Dittrich, and J.-A. Quiané-Ruiz. Runtime Measurements in the Cloud: Observing, Analyzing, and Reducing Variance. *PVLDB*, 3:460–471, 2010.
[46] StratoSphere: Above the Clouds. http://www.stratosphere.eu/.
[47] Teradata, Inc. http://www.teradata.com/.
[48] V. Thummala and S. Babu. iTuned: a Tool for Configuring and Visualizing Database Parameters. In *Proc. of the SIGMOD Conf.*, pages 1231–1234, 2010.
[49] T. White. *Hadoop: The Definitive Guide. MapReduce for the Cloud*. O'Reilly Media, 2009.
[50] I. H. Witten and E. Frank. *Data Mining: Practical machine learning tools and techniques*. Second edition, 2005.
[51] W. Xu, L. Huang, A. Fox, D. Patterson, and M. Jordan. Online System Problem Detection by Mining Patterns of Console Logs. In *ICDM '09: Proceedings of the 2009 Ninth IEEE International Conference on Data Mining*, pages 588–597, 2009.
[52] Y. Yu, M. Isard, D. Fetterly, M. Budiu, U. Erlingsson, P. K. Gunda, and J. Currey. DryadLINQ: A System for General-Purpose Distributed Data-Parallel Computing Using a High-Level Language. In *Proc. of the 8th OSDI Symp.*, pages 1–14, 2008.
609